
\font\ninerm=cmr9  
\font\twerm=cmr12
\font\titlefnt=cmbx10 scaled \magstep2
\magnification 1200
\parindent 0pt

\footline={\hfil}
\rightline{\rm SLAC-PUB-6418}
\rightline{\rm January 1994}
\rightline{\rm(T)}
\baselineskip=6mm
 \vfill
\centerline{\titlefnt
  The Vacuum Theta-Angle is Zero}
\centerline{{\titlefnt
               in Non-Abelian Gauge Theories}\footnote*{\ninerm
Work supported by the Department of Energy, contract DE-AC03-76SF00515.}}
\vskip 1truecm
 \centerline{\twerm Valentin V. Khoze}
\vskip .5truecm
\centerline{\it Stanford Linear Accelerator Center}
\centerline{\it Stanford University, Stanford, California 94309}
\vfill
\centerline{ABSTRACT}
\vskip .5truecm
We argue that gauge transformations with arbitrary fractional winding
numbers should be allowed in a non-Abelian gauge
                                  theory. This eliminates the topological
distinction between large and small gauge transformations.
We prove that the states which satisfy the Gauss's law constraint
are strictly invariant under gauge transformations, i.e. no phase
shift occurs. In other words, the vacuum $\theta$-angle is zero.

Our observation does not resolve the strong CP problem.
\vfill
\centerline{Submitted to Physics Letters B}
\vfill

\vfil\eject
\pageno=1
\footline={\hss\tenrm\folio\hss}
\parindent=20pt

It is usually assumed that the only gauge transformations
allowed in Yang-Mills theory
are those with integer winding numbers. This assumption
is based on some topological argument, which we will discuss later,
and on the properties of instantons.
Instanton solutions describe tunneling processes only between
vacua whose winding numbers differ by an integer.
There exist, however, other classes of solutions in Minkowski space
which correspond to classically allowed transitions between the
configurations with arbitrary winding numbers [1].
If after such a transition we do not want to end up in a state missing
from the classification based on integer winding numbers, we should
expand this classification to include states with arbitrary fractional winding
numbers.

In this note we will point out that the boundary conditions appropriate
to this generalization imply that the physical states of the theory
are invariant under all gauge transformations, and that,
correspondingly,
the vacuum $\theta$-angle is zero.
This conclusion is more conceptual than practical; adding a term
$i \bar{\theta}F\tilde{F}$ to the lagrangian produces observable
CP-violating expectation values with the usual $\bar{\theta}$-dependence.

We first
quantize the theory in the Hamiltonian formalism and then
review the traditional argument which leads to the following
gauge transformation law [2] for the {\it physical} states
of the Hilbert space:
$$ {\cal G}^U |\Psi\rangle_{\rm Phys} = e^{-i\nu[U] \theta}|\Psi\rangle_{\rm
Phys}.
\eqno(1)$$
Here $\nu[U]$ is the (integer) winding number of the gauge transformation
$U$ and ${\cal G}^U$ is the operator in the Hilbert space which implements
this gauge transformation. The physical states are those which obey
the Gauss's law constraint.
The $\theta$-angle is the same for all physical states and it is usually said
that
different values of $\theta$ describe different $\theta$-worlds.
Each $\theta$-world can be viewed as a tower built on the corresponding
$\theta$-vacuum.

Next we allow in the theory gauge transformations with arbitrary fractional
winding numbers. This lifts the topological distinction between large ($\nu
\neq
0$)
and small ($\nu = 0$)
gauge transformations, so that all gauge transformations (large and small)
can be continuously deformed into each other.
Then we prove that under any gauge transformation a physical state transforms
into itself,
$$ {\cal G}^U |\Psi\rangle_{\rm Phys} = |\Psi\rangle_{\rm Phys}.
\eqno(2)$$
This implies that the $\theta$-angle of eq. (1) is always equal to zero.

\vfil\eject

We consider a theory with the action
$$ S= {1\over 2 g^2} \int d^4 x {\rm tr}(F^{\mu \nu} F_{\mu \nu})
,\eqno(3)$$
where the field strength is
$ F_{\mu \nu} = \partial_{\mu} A_{\nu} -\partial_{\nu} A_{\mu}
  + [A_{\mu},A_{\nu}] $
and
$ A_{\mu}=g A_{\mu}^a \tau^a/(2i) $
is the antihermitian vector-potential.
To simplify our presentation we suppress the fermion fields and choose
the number of colors to be two.
The theory is invariant under local gauge transformations
$$ A_{\mu} \to A_{\mu}^U = U^{\dagger}(\partial_{\mu} +A_{\mu})U
,\eqno(4)$$
where $U$ is an $SU(2)$ matrix.

{}From now on we choose the
$A_0 =0$ gauge.
We write the equations of motion,
$D^{\mu} F_{\mu \nu} = 0$
in a non-covariant form,
$$ (D_i E_i)^a =0     ,\eqno(5.a)$$
$$ \partial_t E_i^a = (D\times B)_i^a   ,\eqno(5.b)$$
where $ E_i^a = F_{0i}^a $ and $ B_i^a = -1/2 \epsilon^{ijk} F_{jk}^a$
stand for electric and magnetic fields and
D is the covariant derivative in the adjoint representation,
$(D_i M)^a = \partial_i M^a +g \epsilon^{abc} A_i^b M^c$.
The dynamical variable of the theory is $A_i^a$, the canonical momentum
is $-E_i^a$ and the Hamiltonian is given by
$$ H={1\over2} \int d^3 x (E_i^{a \ 2} +B_i^{a \ 2})
.\eqno(6)$$
The canonical commutation relation,
$$ [ E_i^a({\bf x}),A_j^b({\bf y})]
   =i\delta^{ij} \delta_{ab} \delta^{(3)}({\bf x}-{\bf y})
,\eqno(7)$$
implies
$$ \partial_t A_i^a =i[H,A_i^a] = -E_i^a
,\eqno(8.a)$$
$$ \partial_t E_i^a =i[H,E_i^a] = (D \times B)_i^a
,\eqno(8.b)$$
which reproduces the Amp{\`e}re law (5.b) as an operator equation.
The Gauss law (5.a) should be implemented as a constraint on the
{\it physical} states of the Hilbert space,
$$ (D_i E_i)^a |\Psi\rangle_{\rm Phys}=0     .\eqno(9)$$

A gauge transformation (4) with
$$U({\bf x}) =\exp [\Lambda^a({\bf x}) \tau^a/(2i)]
,\eqno(10)$$
is represented in the Hilbert space by the operator
$$ {\cal G}^U = exp [-i/g \ \int d^3 x E_i^a (D_i \Lambda)^a]
.\eqno(11)$$
Indeed, it is straightforward to check that the following equations
$$ {\cal G}^U A_i {{\cal G}^U}^{\dagger} =U^{\dagger}(\partial_i +A_i)U
,\eqno(12.a)$$
$$ {\cal G}^U E_i {{\cal G}^U}^{\dagger} = U^{\dagger} \ E_i \ U
,\eqno(12.b)$$
with $U$ given by (10), are satisfied by $ {\cal G}^U $ of eq. (11)
purely on the algebraic level, without requiring any integrations
by parts.
We also note that eqs. (12) imply that
$ [{\cal G}^U ,H] =0$.

Thus, eqs. (12) ensure that any gauge transformation which can be cast in
the form (10) is generated by the operator (11).
However, in the standard treatment
[2], the so-called large gauge
transformations cannot be implemented by the exponential operator (11).
Let us review the argument of Refs. [2].
First, one assumes that as ${\bf x} \to \infty$, any gauge function
$U({\bf x})$ approaches a direction-independent matrix.
Then one can compactify the three-space into a three-sphere, $S^3$,
and $U$ provides a map,
$U \ : \ S^3 \rightarrow SU(2) \approx S^3$.
Such maps fall into disjoint homotopy classes labeled by the integers,
called the winding numbers $\nu$,
$$\nu[U] =-{1 \over 24 \pi^2} \int d^3 x \epsilon^{ijk} \ {\rm tr} \bigl( \
(U^{\dagger}\partial_i U)(U^{\dagger}\partial_j U)(U^{\dagger}\partial_k U)
 \ \bigr)
.\eqno(13)$$
Gauge functions belonging to different classes cannot be deformed continuously
into each other
without violating the compactification assumption.
In particular, only those in the zero winding number class
are deformable to the identity and can be generated by the exponential
operator (11). Such gauge transformations are called small and they can
be written in the form (10) with $\Lambda^a ({\bf x}) \to 0$
at spatial infinity. For small gauge transformations the right hand side
of eq. (11) can be integrated by parts and the Gauss's law constraint
(9) requires
$$ {\cal G}^{U_{\rm SMALL}} |\Psi\rangle_{\rm Phys} = |\Psi\rangle_{\rm Phys}.
\eqno(14)$$
But homotopically non-trivial gauge transformations,
according to Refs. [2], are only phase-invariant,
$$ {\cal G}^U |\Psi\rangle_{\rm Phys} = e^{-i\nu[U] \theta}|\Psi\rangle_{\rm
Phys}.
\eqno(15)$$
In this treatment $\theta$ is not constructed, but only
allowed by the disjoint nature of gauge transformations.

We may choose not to impose the compactification assumption and to
allow $U$ to depend on the polar angles at spatial infinity.
There is no difficulty with the boundary conditions; we can,
for example, impose MIT bag [3] boundary conditions,
which contain only $F_{\mu \nu}$ and not $A_{\mu}$ on the
sphere of large radius. In this case,
the above topological classification fails and all gauge functions can
be continuously deformed into each other and are generated by the
exponential operator (11).
As an example of this statement we can consider a gauge function $U$ given
by eq. (10) with
$$ \Lambda^a ({\bf x})= q \Omega (r) \ \hat{x}^a
,\eqno(16)$$
where $q$ is an arbitrary number, $\hat{x}$ is the unit three-vector
in the radial direction and
$$ \Omega(r)|_{r\to 0} \to 0, \ \ \ \ \ \Omega(r)|_{r\to \infty} \to 2\pi
.\eqno(17)$$
The winding number (13) of this gauge function is
$$\nu[U]= q-{\sin(2\pi q)\over 2\pi}
.\eqno(18)$$
Gauge functions of this form can be viewed as a continuous path
of gauge functions labeled by a parameter $q$. This path connects
a gauge function of winding number zero at $q=0$ with a gauge
function of winding number one at $q=1$.
The path interpolates through the gauge functions with fractional
winding numbers at fractional values of $q$.
The operator generating these gauge transformations is given by
eq. (11) with $\Lambda$ and $\Omega$ from eqs. (16)-(17).

Equation (18) illustrates that the definition (13) of the winding
number is somewhat arbitrary. Indeed one might pick
$\nu[U]$ to be any functional of $U$ which coincides with the right hand side
of eq. (13) when $U$ is restricted to a direction-independent matrix
at spatial infinity.

Furthermore we note that eq. (15) with $\nu$ defined by eq. (13) is
somewhat subtle for gauge transformations of this more general type.
Indeed equation (18) implies that for $U_1$ and $U_2$ from the class
(10), (16)-(17) with fractional $q_1$ and $q_2$,
$$ \nu[U_1 U_2] \neq \nu[U_1] +\nu[U_2]
,\eqno(19)$$
and eq. (15) fails to represent the group.
Of course, one may think that
what we have here is a group representation up to a phase
or {\it projective} representation:
$$ {\cal G}^{U_1} {\cal G}^{U_2} \ |\Psi\rangle_{\rm Phys}=
 \eqno(20) $$
$$
   \exp\bigl\lbrack \ i\theta /(2\pi) \ \{ \ \sin(2\pi(q_1+q_2)) -
   \sin(2\pi q_1) - \sin(2\pi q_2) \ \} \
   \bigr\rbrack \ {\cal G}^{U_1 U_2} \
|\Psi\rangle_{\rm Phys},
$$
with
$$
{\cal G}^{U_j}= \exp [-iq_j/g \int d^3 x \ E_i^a \ (D_i \Omega (r) \
\hat{x})^a]
.\eqno(21)$$
But now using eq. (21) we can immediately see that
$${\cal G}^{U_1}
\ {\cal G}^{U_2} \ = \
 \exp [-i(q_1+q_2)/g \int d^3 x \ E_i^a \ (D_i \Omega (r) \ \hat{x})^a]
\equiv {\cal G}^{U_1 U_2} ,$$
which is in a contradiction with eq. (20).
This demonstrates that eq. (15) with $\nu$ defined by eq. (13)
is not a valid statement for gauge transformations of the type (10), (16)-(17).
Even more generally, eq. (15) fails for any gauge transformation
which can be transformed to a gauge transformation of the class
(10), (16)-(17) making a small gauge transformation.

The only way to have no phase in (20) is to appropriately modify the definition
of the winding number (13) for fractional values of $\nu$, that is
we must choose $\nu=q$ instead of (13), and the gauge transformation law (15)
then takes the form,
$$ {\cal G}^{U(q)} |\Psi\rangle_{\rm Phys} = e^{-iq \theta}|\Psi\rangle_{\rm
Phys}
.\eqno(22)$$
For integer values of $q$, eqs. (22) and (15) certainly agree.

Now we will prove that the $\theta$-angle on the right hand side of eq. (22)
is zero.
The left hand side of equation (22) should be viewed as an expansion
in powers of $q$ and when we equate the terms proportional
to the $q^1$ on the right and on the left we obtain,
$$ \theta |\Psi\rangle_{\rm Phys}= 1/g \int d^3 x  \ E_i^a
            \bigl(D_i \ \Omega (r) \hat{x} \bigr)^a \  |\Psi\rangle_{\rm Phys}
$$

$$ =1/g \int d^3 x \ \biggl(\partial_i(E_i^a \Omega (r)\hat{x}^a)
    - \Omega (r) \hat{x}^a (D_i E_i)^a \biggr) \ |\Psi\rangle_{\rm Phys}
 \eqno(23)$$
$$   =2\pi /g \int d^3 x \ \partial_i\bigl(\hat{x}^a  E_i^a \bigr)
|\Psi\rangle_{\rm Phys}
,$$
 where the Gauss's law constraint (9) was used.
We note that if we have restricted ourselves to MIT bag boundary
conditions [3], this would imply that the component of $E_i^a$
normal to the boundary
is zero at the boundary, and the right hand
side of eq. (23) is zero. This implies that $\theta$ is zero for MIT
bag
boundary conditions.
We will demonstrate that this is also true
in general.
Let us
gauge transform the left and the right hand side of eq. (23) with
the gauge function $V$ of the form,
$$V({\bf x}) =\exp [ \Phi (r) \ n^a\tau^a/(2i)]
,\eqno(24)$$
where $\Phi$ is some function of $r$ and $n^a$ is an arbitrary
$x$-independent unit-vector.
The gauge transformation of eq. (23) reads,
$$ \theta \ {\cal G}^V |\Psi\rangle_{\rm Phys}=1/g \int d^3 x \ \partial_i
   \bigl(\Omega (r) \hat{x}^a \ {\cal G}^V  E_i^a {{\cal G}^V}^{\dagger}
    \bigr) {\cal G}^V |\Psi\rangle_{\rm Phys}
.$$
Under the action of ${\cal G}^V$ the state $ |\Psi\rangle_{\rm Phys}$
can at most acquire
a phase which we cancel on the left and on the right
of the above equation and arrive at the following
expression,
$$ \theta |\Psi\rangle_{\rm Phys}=1/g \int d^3 x \ \partial_i
   \bigl(\Omega (r) \hat{x}^a \ {\cal G}^V  E_i^a {{\cal G}^V}^{\dagger}
    \bigr) |\Psi\rangle_{\rm Phys}
,\eqno(25)$$
where the gauge transformation of the electric field reads in this
case,
$$ {\cal G}^V  E_i^a {{\cal G}^V}^{\dagger}=
   E_i^a \cos(\Phi) +\epsilon^{abc}E_i^b n^c \sin(\Phi)
   +2 n^a n^b E_i^b \sin^2(\Phi/2)
.\eqno(26)$$
Equation (25) should hold for any $V$ of the form (24), in other words the left
hand
side  of eq. (25) is  independent of asymptotic value of $\Phi$ and on the
choice of
$n^a$.
Let us choose $\Phi$ such that $\Phi |_{r\to \infty} \to \pi/2 +\alpha$,
differentiate both sides of eq. (26) with respect to $\alpha$ and then
set  $\alpha$
equal to zero.
This establishes the following equality:
$$  2\pi/g \int d^3 x \ \partial_i\bigl(\hat{x}^a  E_i^a
     -\hat{x}^a n^a   E_i^b n^b \bigr)
    |\Psi\rangle_{\rm Phys} =0
,\eqno(27)$$
which is possible only if
$$ \int d^3 x \ \partial_i\bigl(\hat{x}^a  E_i^a \bigr)
 \ |\Psi\rangle_{\rm Phys} =0
.\eqno(28)$$
This proves that $\theta=0$ and that the
transformation law
$$ {\cal G}^U |\Psi\rangle_{\rm Phys} = |\Psi\rangle_{\rm Phys}
\eqno(29)$$
holds for arbitrary gauge transformations.

We mention that the derivation of eq. (28) is based on the fact that
$E_i^a$ is a gauge-variant operator. Thus our conclusion, eq. (29)
is not necessarily true in Abelian models, see Ref. [4].

The gauge transformation law (29) implies that the so-called
theta-vacua do not contain any theta and thus the vacuum is unique.
It is given just by the integral over the whole gauge orbit of
the perturbative vacuum.
We note again that when fractional winding numbers are allowed
there is just one gauge orbit, that is all gauge transformations
can be continuously deformed to each other.

It is worthwhile mentioning that the $\theta F \tilde{ F}$ term
can be obtained in the functional integral formalism
starting from the $\theta$-independent classical action, eq. (3),
and the $\theta$-dependent gauge transformation, eq. (15)
[2]. On the other hand we have just argued that $\theta=0$
in eq. (15). This observation, however, does not  prove
that the $\theta F \tilde{ F}$ term is zero.
In fact, one can just start with the action which contained
$\bar{\theta} F \tilde{ F}$.
The canonical momentum, $\Pi_i^a$,
of this theory is now $\bar{\theta}$-dependent.
The gauge transformation operator, however, remains (11)
with $E_i^a$ and not $-\Pi_i^a$ in the exponent.
This demonstrates that $\bar{\theta}$ cannot be removed from the formalism.
Thus the $\bar{\theta}$-strong CP parameter does not seem to have much in
common
with the $\theta$-vacuum-angle which we proved is zero.

\vskip 1truecm
\centerline{\bf Acknowledgments}
We are very grateful to Michael Peskin,
Helen Quinn and Marvin Weinstein for most valuable
discussions and comments. We also thank Arnd Brandenburg, Edward Farhi,
Alex Kagan, Eric Sather and Peter van Driel
for useful conversations and comments
on the manuscript.

\vfil\eject

\centerline{\bf References}
\item{[1]}
E. Farhi, V.V. Khoze and R. Singleton, {\it Phys. Rev.} {\bf D 47}, 5551
(1993).

\item{[2]}
R. Jackiw and C. Rebbi, {\it Phys. Rev. Lett.} {\bf 37}, 172 (1976);

\item{}
C. Callan, R. Dashen and D. Gross, {\it Phys. Lett.} {\bf 63 B}, 334 (1976).

\item{[3]} A. Chodos, R.L. Jaffe, K. Johnson and V.F. Weisskopf,
{\it Phys. Rev.} {\bf D 12}, 3471 (1974).

\item{[4]}
H.R. Quinn and M. Weinstein, {\it Phys. Rev.} {\bf D 17}, 1063 (1978).

\bye